\documentclass[aps,prl,twocolumn,floatfix]{revtex4}

\usepackage{graphicx}

\def\gtorder{\mathrel{\raise.3ex\hbox{$>$}\mkern-14mu
 \lower0.6ex\hbox{$\sim$}}}
\def\ltorder{\mathrel{\raise.3ex\hbox{$<$}\mkern-14mu
 \lower0.6ex\hbox{$\sim$}}}
\def\mugegm{\mu_p G_E / G_M}

\def\gep{G_E}
\def\gmp{G_M}
\def\etal{\textit{et al.}}

\begin{document}

\title{Comment on ``High-Precision Determination of the Electric and Magnetic
Form Factors of the Proton''}

\author{J. Arrington}

\affiliation{Physics Division, Argonne National Laboratory, Argonne, Illinois 60439, USA}

\date{\today}

\maketitle


In a recent Letter, Bernauer, \etal~\cite{bernauer10} present fits to the
proton electromagnetic form factors, $\gep(Q^2)$ and $\gmp(Q^2)$, along with
extracted proton charge and magnetization radii based on large set of new,
high statistical precision ($<$0.2\%) cross section measurements.  The
Coulomb corrections (CC) they apply~\cite{mckinley48} differ dramatically
from more modern and complete calculations, implying significant error in their
final results.

\begin{figure}[htb]
\includegraphics[width=7.5cm,angle=0]{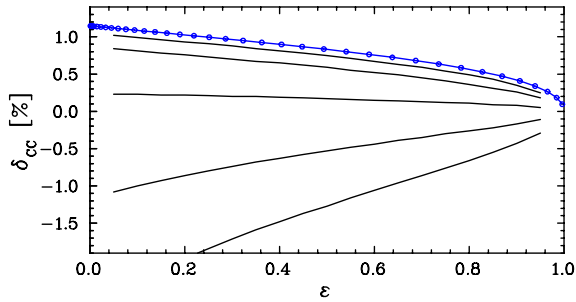}
\caption{(Color online) The Coulomb correction from Ref.~\cite{mckinley48}
(circles), evaluated at the mean $Q^2$ of the experiment, and the full CC
result~\cite{arrington04c} for $Q^2$=0.01 (top), 0.03, 0.1, 0.3, and
1.0~GeV$^2$ (bottom).  TPE calculations~\cite{blunden05a, borisyuk06b}
yield similar results, with a somewhat weaker $Q^2$ dependence at low
$Q^2$.
\label{fig:compare}}
\end{figure}

It has been shown that two-photon exchange (TPE) corrections are important
in the extraction of the form factors~\cite{arrington07c} and the
charge radius~\cite{blunden05b} of the proton.  At low $Q^2$, the Coulomb
correction (representing the soft part of the TPE) yields the dominant
contribution and has a significant $Q^2$ dependence at very low
$Q^2$~\cite{arrington04c, blunden05a, borisyuk06b}. In the analysis of
Ref.~\cite{bernauer10}, the applied correction~\cite{mckinley48, bernauerphd}
is the $Q^2 \to 0$ limit of the full calculation:
\begin{equation}
\delta_{CC} = Z \alpha \pi [\sin(\theta/2)-\sin^2(\theta/2)]/\cos^2(\theta/2) ~.
\label{eq:delta}
\end{equation}
Figure~\ref{fig:compare} shows the CC applied in Ref.~\cite{bernauer10}
along with the full $Q^2$-dependent result~\cite{arrington04c}. The full
correction is outside of the 50\% uncertainty assumed in
Ref.~\cite{bernauer10} for all data above $Q^2=0.06$~GeV$^2$. By
0.1~GeV$^2$, the small-$\varepsilon$ correction has changed by 1\% which will
modify $\gmp$ and its $Q^2$ dependence, altering the extracted magnetic radius.
The full $\delta_{CC}$ is 2--3\% lower for $Q^2>0.3$~GeV$^2$ and low
$\varepsilon$: a change several times the the total uncertainties on the
individual cross sections (which do not include any systematic uncertainties,
although all kinematic settings have inflated statistical errors to account
for non-statistical deviations from the global fit~\cite{bernauerphd}).  The
fits include estimates of systematics and theoretical (TPE) uncertainties
which are essentially negligible at small scattering angles and at most
$\sim$0.5\% at large angles~\cite{bernauerphd}, still much smaller than the
error in $\delta_{CC}$.

\begin{figure}[htb]
\includegraphics[width=7.5cm,angle=0]{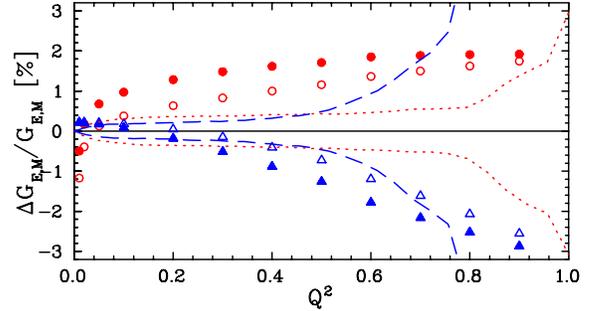}
\caption{(Color online) Estimated change in $\gmp$ (red circles) and $\gep$
(blue triangles).  The points show the impact of replacing the
CC of Ref.~\cite{bernauer10} with the full CC~\cite{arrington04c} (solid 
symbols) or TPE calculation~\cite{blunden05a} (hollow symbols),
using dipole form factors and assuming that the
cross section data cover $0.3 < \varepsilon < 0.9$.  The dotted (dashed)
lines show the fit uncertainties on $\gmp$ ($\gep$)~\cite{bernauer10}. \label{fig:error}}
\end{figure}

Figure~\ref{fig:error} shows the estimated impact of the full CC or TPE
calculations on a direct Rosenbluth separation of the form factors. This
suggests that proper implementation of the corrections will shift the
$\gmp$ results by more than 2--3 standard deviations, bringing the
ratio $\mugegm$ into better agreement with recent high-precision
polarization measurements~\cite{zhan11}.

This work was supported by the U.S. DOE, Office of
Nuclear Physics, under contract DE-AC02-06CH11357.

\bibliography{comment_bernauer}

\end{document}